\documentclass{PoS}

\title{Eta-mesic nuclei}

\ShortTitle{Eta-mesic nuclei}

\author{Ale\v{s} Ciepl\'{y}, \speaker{Ji\v{r}\'{\i} Mare\v{s}}, Martin Sch\"{a}fer\\
        Nuclear Physics Institute, 250 69 Rez, Czech Republic\\
        E-mail: \email{mares@ujf.cas.cz} 
}

\author{Nir Barnea, Betzalel Bazak, Eliahu Friedman, Avraham Gal\\
        Racah Institute, The Hebrew University, Jerusalem 91904, Israel 
}

\abstract{
In this contribution we report on theoretical studies of $\eta$ nuclear quasi-bound states in few- and 
many-body systems performed recently by the Jerusalem-Prague Collaboration~\cite{fgm13,cfgm14,bfg15,cit2,cit3}. 
Underlying energy-dependent $\eta N$ interactions are derived from coupled-channel models
that incorporate the $N^*(1535)$ resonance. The role of self-consistent treatment of the
strong energy dependence of subthreshold $\eta N$ amplitudes is discussed. 
Quite large downward energy shift together with rapid decrease of the $\eta N$ amplitudes below threshold result in relatively 
small binding energies and widths of the calculated $\eta$ nuclear bound states. 
We argue that the subthreshold behavior of $\eta N$ scattering amplitudes is crucial to conclude whether  
$\eta$ nuclear states exist, in which nuclei the $\eta$ meson could be bound and if the corresponding widths are small enough to 
allow detection of these $\eta$ nuclear states in experiment.        
}

\FullConference{XVII International Conference on Hadron Spectroscopy and Structure - Hadron2017\\
		25-29 September, 2017\\
		University of Salamanca, Salamanca, Spain}

\begin{document}

\section{Energy and model dependence of $\eta N$ scattering amplitudes}

Calculations of $\eta$ nuclear quasi-bound states presented in this contribution are 
based on the $\eta N$ scattering amplitudes derived from coupled-channel models that incorporate the  
$N^{\ast}(1535)$ resonance. 
The amplitudes near threshold are both attractive and strongly energy dependent, as illustrated 
in Fig.~1 for three selected meson-baryon interaction models, GW~\cite{gw5}, CS~\cite{cs13}, 
and GR~\cite{gr02}.
Moreover, the $\eta N$ scattering amplitudes are highly model dependent; they differ considerably 
from each other below as well as above the $\eta N$ threshold (except  
common value  Im$F_{\eta N}\approx 0.2-0.3$~fm at threshold).  
This suggests that the predictions for the $\eta$ nuclear states would be model dependent and 
that the strong energy dependence of the $\eta N$ scattering amplitudes  
has to be treated self-consistently.  
\begin{figure}[h!]
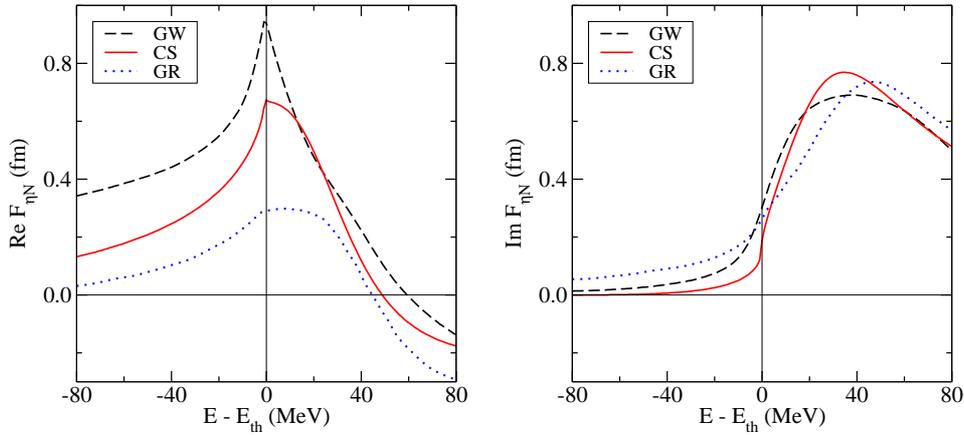
 
\begin{center} 
\includegraphics[width=0.4\textwidth]{retamod.eps} 
\hspace*{8pt}
\includegraphics[width=0.4\textwidth]{ietamod.eps} 
\caption{Real (left panel) and imaginary (right panel) parts of the free $\eta N$ 
c.m. scattering amplitude $F_{\eta N}(\sqrt{s})$ as a function of 
energy in three meson--baryon interaction models: 
dashed, GW~\cite{gw5}; solid, CS~\cite{cs13}; dotted, GR~\cite{gr02}. 
The vertical line denotes the $\eta N$ threshold.} 
\label{fig:aEtaN1} 
\end{center} 
\end{figure} 

The crucial point is that in the nuclear medium 
the energy argument $\sqrt{s}$  is given by  
\begin{equation}
\sqrt{s}= \sqrt{(\sqrt{s_{\rm th}}-B_{\eta}-B_N)^2 -(\vec{p}_{\eta} +\vec{p}_N)^2} \leq \sqrt{s_{\rm th}},
\end{equation}
where $\sqrt{s_{\rm th}}\equiv m_{h}+m_N$ and $B_{\eta}$ and $B_N$ are meson and nucleon 
binding energies, and the momentum dependent term generates additional substantial downward energy shift, since 
 $(\vec{p}_{\eta} +\vec{p}_N)^2 \neq 0$  unlike the case of the two-body c.m. system.  
This has significant consequences for the calculated binding energies and widths as will be shown below.  


\section{The $\eta$ meson in few-body systems}
Few-body calculations of $\eta$ nuclear clusters have been performed within standard few-body techniques: 
Faddeev-Yakubovsky equations~\cite{fix17} 
or variational methods. In ref.~\cite{bfg15} the $\eta$ nuclear  cluster wave functions were expanded in a hyperspherical basis. 
More recent calculations~\cite{cit2,cit3} were based on the Stochastic Variational Method (SVM) with a correlated Gaussian 
basis~\cite{cit1}. Both variational approaches showed sufficient accuracy in the description of $\eta$ nuclear quasi-bound states
and provided almost identical results for $\eta d$, $\rm \eta ^3He$ and $\rm \eta ^4He$ systems.  

In our calculations, the nuclear part is described by the Minnesota central $NN$ potential~\cite{mn} or the Argonne AV4' 
potential~\cite{arg}. The interaction of $\eta$ with nucleons of the core is given by a complex two-body energy dependent potential 
derived from a full chiral coupled-channels model:
\begin{equation} 
v_{\eta N}(\delta\sqrt{s}, r) = - \frac{4\pi}{2\mu_{\eta N}} b(\delta\sqrt{s}) \rho_{\Lambda}(r), 
\end{equation}    
 where $\delta\sqrt{s} = \sqrt{s} - \sqrt{s_{\rm th}}$, \quad $\rho_{\Lambda}(r) = (\frac{\Lambda}{2\sqrt{\pi}})^3 exp\left(\ -\frac{\Lambda^2 r^2}{4}\right)$, and the amplitude $b(\delta\sqrt{s})$ is fitted to phase shifts derived from 
the $\eta N$ scattering amplitude  $F_{\eta N}(\delta\sqrt{s})$ in the GW and CS models. 
The scale parameter $\Lambda$ is inversely proportional to the range of $V_{\eta N}$ potential. We consider two different values 
of the scale parameter, $\Lambda =2$ and 4 fm$^{-1}$ (the choice of the value of $\Lambda$ is discussed in ref.~\cite{bfg15}). 
It is to be noted that in ref.~\cite{cit3}, the $NN$ and $\eta N$ potentials were constructed within a pionless EFT approach. 

The energy argument $\delta\sqrt{s}$ relevant for calculations of $\eta$ nuclear few-body clusters is expressed in the 
form~\cite{bfg15}: 
\begin{equation}
\delta\sqrt{s} =  - \frac{B}{A} - \frac{A-1}{A} B_{\eta}  - \xi_N \frac{A-1}{A}\langle T_{NN} \rangle  
- \xi_{\eta} \left(\frac{A-1}{A}\right)^2 \langle T_{\eta} \rangle\; ,
\end{equation}
where $B$ is the total binding energy of the system, 
 $\xi_{N(\eta)} = m_{N(\eta)}/(m_N+m_{\eta})$,
$T_{\eta}$ is the $\eta$ kinetic energy  
in the total c.m. frame and $T_{NN}$ is the pairwise $NN$ kinetic energy operator in the $NN$ pair 
c.m. system~\cite{bfg15}. 
 The conversion widths are calculated using the expression 
\begin{equation}
\Gamma_{\eta}=-2<\Psi_{g.s.}|{\rm Im} V_{\eta N}|\Psi_{g.s.}>
\end{equation}
 where $|\Psi_{g.s.}>$ stands for the ground state obtained after variation. As was stated already in~\cite{bfg15}, 
this approximation 
 is reasonable due to small imaginary contribution $|{\rm Im} V_{\eta N}| \ll |{\rm Re} V_{\eta N}|$.

The results of calculations of $\eta$ nuclei with $A=3$ and 4 were discussed in detail in refs.~\cite{bfg15, cit2, cit3}. 
To summarize, no bound $\eta NN$ system was found in the considered two-body interaction models.  
For $\eta NNN$, a weakly bound state (with $\eta$ separation energy below 1 MeV) 
was found for the Minnesota $NN$ potential and one particular variant of the $\eta N$ potential that 
reproduced the GW scattering amplitudes. No $\eta NNN$ bound states were 
found using more realistic $NN$ interaction model.

\begin{figure}[h!]
\centering
\includegraphics[width=0.6\textwidth]{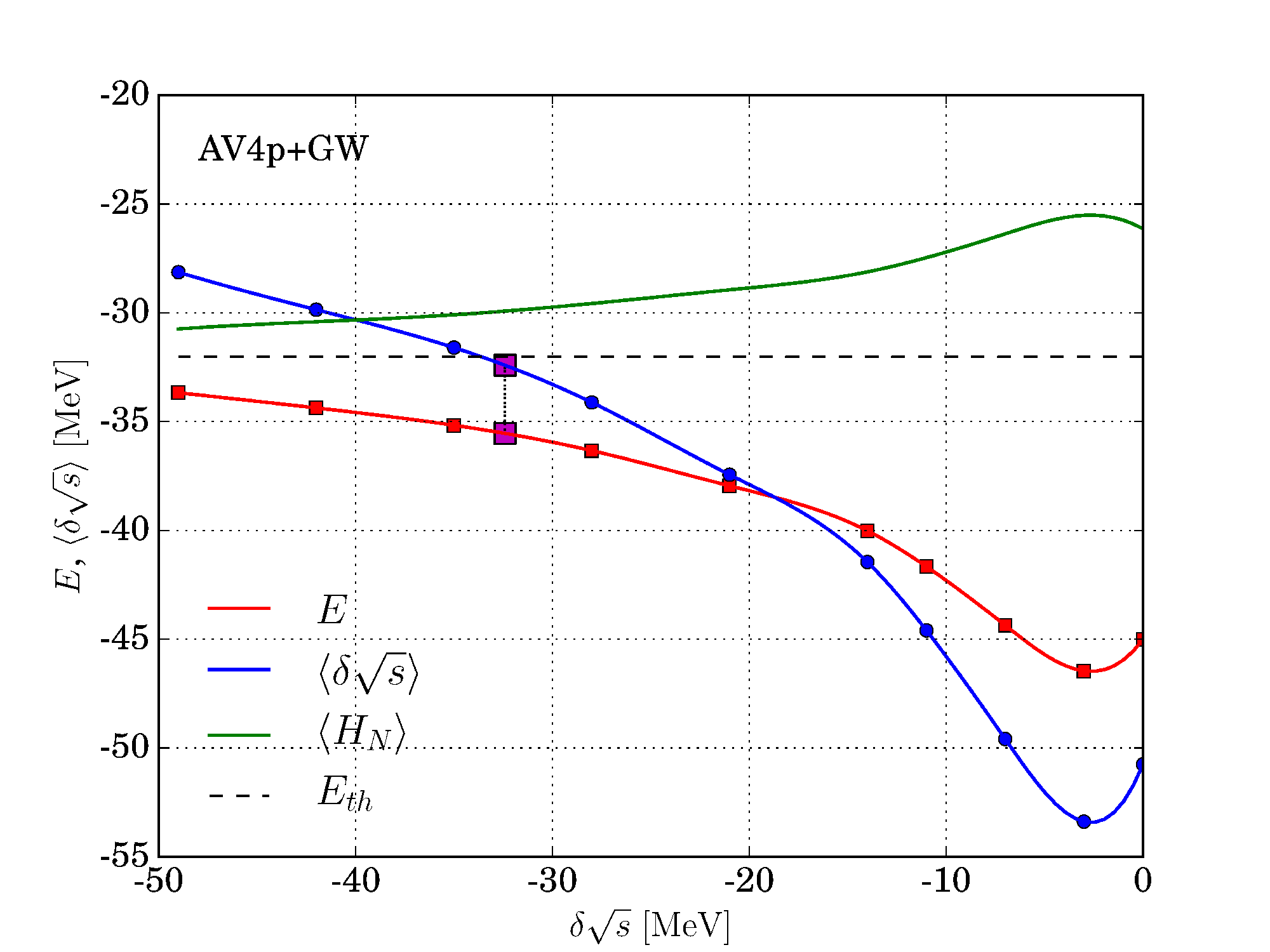}
\caption{$\eta ^4$He bound state energy E (red line, squares) and the expectation value 
$\langle \delta \sqrt{s} \rangle$ (blue line, circles), calculated using the AV4' $NN$ potential 
(denoted here AV4p), as a function of the input energy argument $\delta\sqrt{s}$ of the 
$\eta N$ potential GW with $\Lambda=4$~fm$^{-1}$. The dotted vertical line marks the 
self-consistent output values of $\langle \delta\sqrt{s}\rangle$ and $E$. The black dashed line 
denotes the $^4$He g.s. energy which serves as threshold for bound $\eta$. The green curve shows 
the expectation value $<H_N>$ of the nuclear core energy. Figure adapted from ref.~\cite{cit2}.   
}  
\label{pic1}
\end{figure}
In Fig. \ref{pic1}, we demonstrate the self-consistent solution for $\eta ^4$He, calculated using 
the AV4' $NN$ potential and GW $V_{\eta N}$ potential with $\Lambda =4~ {\rm fm^{-1}}$. 
The $\eta ^4$He bound state energy $E$ and the expectation value $<\delta \sqrt{s}>$ are plotted as a function of 
the subthreshold energy argument $\delta \sqrt{s}$ of the input potential $V_{\eta N}$. The self-consistency 
condition is fulfilled by requiring $ \delta \sqrt{s} = <\delta \sqrt{s}>$. The corresponding value of 
$E(<\delta \sqrt{s}>)$  then represents the self-consistent energy of the $\eta$ nuclear cluster.

A precise self-consistent calculation of $p$-shell $\eta$ nuclear clusters, such as $\rm \eta ^6Li$, 
represents highly non-trivial goal. 
In this report, we present our preliminary results for $\rm \eta ^6 Li$ using the central Minnesota $V_{NN}$ and GW $V_{\eta N}$ 
potentials. This should be regarded as the first step before doing calculations with a more realistic $NN$ 
potential to account for spin dependent force components in the $p$ shell. Moreover,      
we employed only one spin-isospin configuration in the description of the 
$\rm ^6Li$ nuclear core, which yielded binding energy $B({\rm ^6 Li})=34.66$ MeV. 
It is reasonable to expect that taking into account all possible 
configurations in $^6$Li will further increase the binding.~\footnote
{In ref.~\cite{nc14}, a value of $B({\rm ^6 Li})=36.51$~MeV was quoted for the SVM calculation with the Minnesota 
potential when more spin-isospin configurations were considered.} 
A full account will be given elsewhere in due course.

\begin{figure}[h!]
\centering
\includegraphics[width=0.8\textwidth]{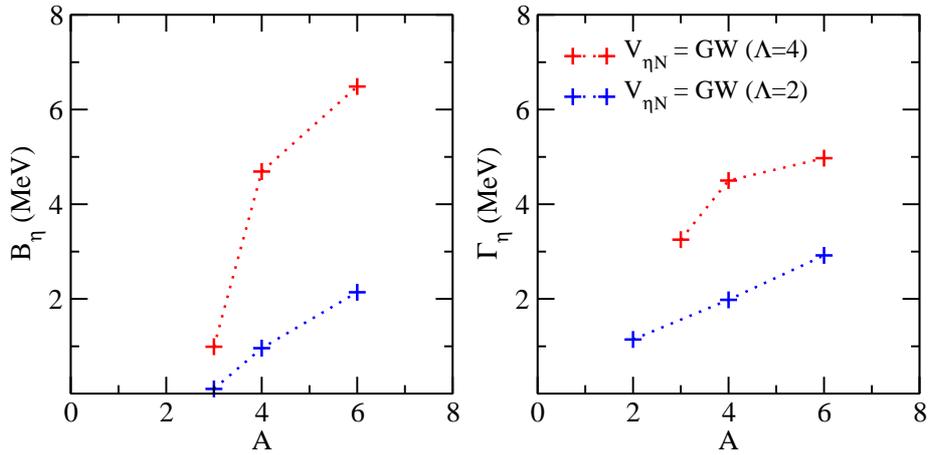}
\caption{
Binding energies $B_{\eta}$ (left) and widths $\Gamma_{\eta}$ (right) of $1s\; \eta$ quasi-bound states in few-body nuclear systems calcualted using the Minnesota $NN$ potential and the $\eta N$ potential GW with 
$\Lambda = 2$ and 4~fm$^{-1}$.}  
\label{pic2}
\end{figure}

The results of the SVM calculations of $\eta$ binding energies $B_\eta$ and widths $\Gamma_\eta$ in $\rm \eta ^3H$, 
$\rm \eta ^4He$, and $\rm \eta ^6Li$ are summarized in Fig. \ref{pic2}. Moreover, the figure illustrates the extent of the dependence of $B_\eta$ and $\Gamma_\eta$ on the parameter $\Lambda$.

\section{The $\eta$ meson in many-body systems}

The binding energies $B_{\eta}$ and widths $\Gamma_{\eta}$ of $\eta$ quasi-bound states in nuclear many-body systems 
are determined by solving self-consistently the Klein-Gordon equation 

\begin{equation} 
[\:\nabla^2+{\tilde\omega}_{\eta}^2-m_{\eta}^2-\Pi_{\eta}(\omega_{\eta},\rho)\:]\:\psi=0  
\; , 
\label{eq:kg} 
\end{equation} 
where ${\tilde\omega}_{\eta}= \omega_{\eta} -{\rm i}\Gamma_{\eta}/2$ is complex energy of $\eta$, 
$\omega_{\eta} = m_{\eta} -B_{\eta}$. 
The self-energy operator 
$
\Pi_{\eta}(\sqrt{s},\rho)\equiv 2\omega_{\eta}V_{\eta}=-({\sqrt{s}}/{E_N})\,4\pi
F_{\eta N}(\sqrt{s},\rho)\rho
$  
is constructed self-consistently using the relevant in-medium $\eta N$ scattering amplitude $F_{\eta N}(\sqrt{s}$) 
and RMF density of the core nucleus.

Modifications of the free-space amplitudes GW due to Pauli blocking in the medium are accounted for by using 
the multiple scattering approach~\cite{wrw}. In the chirally inspired meson-baryon interaction models 
CS and GR, Pauli blocking restricts integration domain in the in-medium Green's function  
which enters the underlying Lippmann-Schwinger (Bethe-Salpeter) equations~\cite{cs13}. Morever, 
hadron self-energy insertions reflecting in-medium modifications of hadron masses could be included in the 
in-medium Green's function, as well. 

The energy argument in the scattering amplitude $F_{\eta N}(\sqrt{s})$ is approximated as~\cite{fgm13} 
\begin{equation} 
\delta\sqrt{s} = \sqrt{s} - \sqrt{s_{\rm th}} \approx -B_N\frac{\rho}{{\bar\rho}}-
\xi_N B_{\eta}\frac{\rho}{\rho_0}-\xi_N T_N(\frac{\rho}{\rho_0})^{2/3}-
\xi_{\eta} \frac{\sqrt{s}}{\omega_{\eta} E_N}2\pi{\rm Re}~F_{\eta N}(\sqrt{s},\rho)\rho \; ,
\label{eq:sqrts}
\end{equation}
where $\bar\rho$ is the average nuclear 
density, $T_N=23.0$ MeV at $\rho_0$, and $B_N\approx 8.5$~MeV is the average nucleon binding energy.
It is to be stressed that all terms in Eq.~\ref{eq:sqrts} are negative definite and thus provide   
substantial downward energy shift. 
Since ${\rm Re}F_{\eta N}(\sqrt{s})$ and  $B_{\eta}$ appear 
as arguments in the expression for $\delta\sqrt{s}$ (Eq.~\ref{eq:sqrts}), which in turn serves as 
an argument for the self-energy $\Pi_{\eta}$ in Eq.~\ref{eq:kg}, a self-consistency scheme 
is required in calculations.~\footnote{A slightly different form of $\delta\sqrt{s}$ 
has been used in recent calculations~\cite{fg17,hm17}, see the contribution of A. Gal in these proceedings.}  

\begin{figure}[b!]
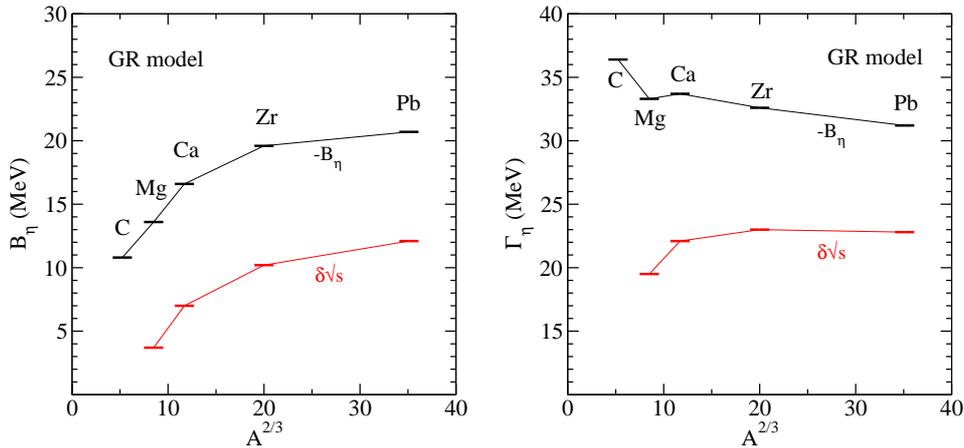
 
\begin{center} 
\includegraphics[width=0.4\textwidth]{beta-ales-oset.eps} 
\hspace{8pt} 
\includegraphics[width=0.4\textwidth]{gamma-ales-oset.eps} 
\caption{Binding energies (left) and widths (right) of the $1s$ ${\eta}$ nuclear 
states in selected nuclei calculated using the GR $\eta N$ scattering amplitude \cite{gr02}  
with different procedures for subthreshold energy shift $\delta\sqrt{s}$.} 
\label{fig:ales-oset} 
\end{center} 
\end{figure} 
It is instructive to compare our self-consistency procedure based on $\delta\sqrt{s}$ of Eq.~\ref{eq:sqrts}, 
with a self-consistency requirement $\delta\sqrt{s} = -B_{\eta}$  applied in Ref.~\cite{car02}.  
This comparison is presented in Fig.~\ref{fig:ales-oset} for the in-medium GR amplitude. Our self-consistency 
formula in Eq.~\ref{eq:sqrts} (marked $\delta\sqrt{s}$) reduces considerably binding energies and widths of the $\eta$ meson in nuclei 
with respect to the calculations of ref.~\cite{car02} that used $\delta\sqrt{s}=-B_{\eta}$ (marked $-B_{\eta}$). 
However, even the reduced GR widths are still rather large, which suggests that it would be extremely difficult 
to resolve $\eta$ nuclear states in this case.  

\begin{figure}[t!]
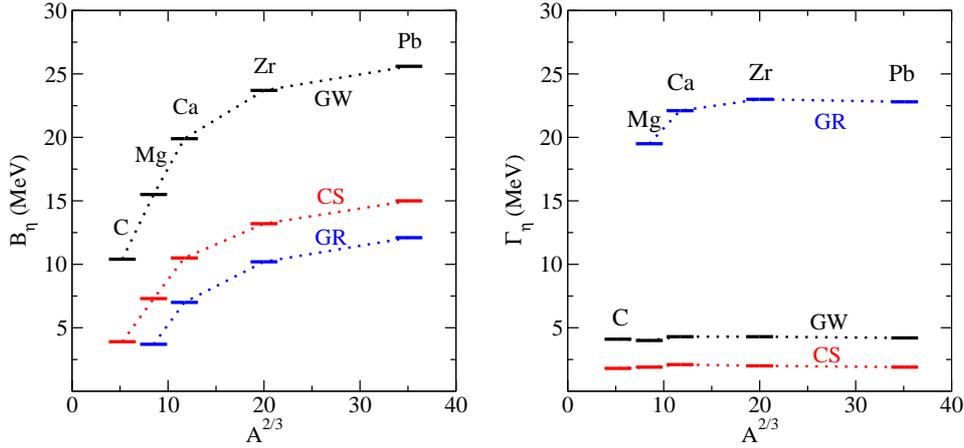
 
\begin{center} 
\includegraphics[width=0.4\textwidth]{beta-all.eps} 
\hspace{8pt} 
\includegraphics[width=0.4\textwidth]{gamma-se2.eps} 
\caption{Binding energies (left) and widths (right) of $1s$ ${\eta}$ nuclear 
states in selected nuclei accross the periodic table calculated self consistently using the 
GW, GR, and GR $\eta N$ scattering amplitudes.} 
\label{fig:bg-s} 
\end{center} 
\end{figure}

The model dependence of the $\eta N$ amplitudes, shown in Fig.~1, has an impact on the 
calculations of $\eta$ nuclear quasi-bound states. This is 
illustrated in Fig.~\ref{fig:bg-s} were we present binding energies $B_{\eta}$ and widths $\Gamma_{\eta}$ 
calculated for $1s$ ${\eta}$ 
nuclear states in selected nuclei using the GW, CS and GR models. 
In the left panel, the hierarchy of the three curves for the $\eta$ binding energies reflects the 
strength of the Re$F_{\eta N}(\sqrt{s})$ amplitudes below threshold (compare Fig.~1). For each $\eta N$ 
interaction model the binding energy increases with $A$ and tends to saturate for large values of $A$. 

The right panel demonstrates substantial differences between the $\eta$ absorption 
widths $\Gamma_{\eta}$. While the CS and GW models    
produce relatively small widths (2 to 4 MeV), almost constant across the periodic table,  
the GR model yields much larger widths  of order 20~MeV which increase with $A$.
 

\section{Conclusions}
In this contribution we briefly reviewed our calculations of $\eta$ nuclear quasi-bound states accross the 
periodic table. We applied $\eta N$ scattering amplitudes derived from recent meson-baryon coupled-channel 
interaction models. We demonstrated that the strong energy dependence of scattering amplitudes calls for proper 
self-consistent treatment. The corresponding $\eta N$ amplitudes relevant for calculations of $\eta$ nuclear 
states are substantially weaker than the $\eta N$ scattering lengths. As a result our calculated $\eta$ bound 
states energies and widths are considerably smaller than those obtained in other comparable calculations. 
  
In few-body calculations we explored whether the $\eta$ meson binds in light nuclei.  
We found no $\eta NN$ bound state. Our results suggest that the onset of $\eta ^3$He binding 
occurs for the models providing the $\eta N$ scattering length Re$a_{\eta N} \sim 1$~fm.   
The binding $\eta ^4$He requires Re$a_{\eta N} \geq 0.7$~fm. It is to be noted that 
the searches for $\eta ^4$He bound states performed with the WASA-at-COSY facility have not 
revealed any signal for a narrow $\eta$ nuclear state~\cite{wasa}.

Small conversion widths in heavier $\eta$ nuclei obtained in calculations using the CS and GW amplitudes 
might encourage experimental searches for $\eta$ nuclear bound states~\footnote{Additional contributions 
to the widths due to $\eta N \rightarrow \pi\pi N$ and $\eta NN \rightarrow NN$ processes, disregarded in 
our calculations, are estimated to add a few MeV to the total $\eta$ nuclear widths.}   
It is to be stressed, however, that 
the size of the widths $\Gamma_{\eta}$ and binding energies $B_{\eta}$ is strongly model dependent. 
Other models produce either substantially larger widths or even do not generate any $\eta$ nuclear bound 
state in a given nucleus.

\acknowledgments
This work was supported by the GACR Grant No. P203/15/04301S.


\begin{thebibliography}{99}
\bibitem{fgm13} E. Friedman, A. Gal, J. Mare\v{s}, \emph{Phys. Lett.} {\bf B 725} (2013) 334.
\bibitem{cfgm14} A. Ciepl\'{y}, E. Friedman, A. Gal, J. Mare\v{s}, \emph{Nucl. Phys.} {\bf A 925} (2014) 126.
\bibitem{bfg15} N. Barnea, E. Friedman, A. Gal, \emph{Phys. Lett.} {\bf B 747} (2015) 345.
\bibitem{cit2} N. Barnea, E. Friedman, A. Gal, \emph{Nucl. Phys.} {\bf A 968} (2017) 35. 
\bibitem{cit3} N. Barnea, B. Bazak, E. Friedman, A. Gal, \emph{Phys. Lett.} {\bf B 771} (2017) 297. 
\bibitem{gw5} A. M. Green, S. Wycech, \emph{Phys. Rev.} {\bf C 71} (2005) 014001.
\bibitem{cs13}  A. Ciepl\'{y}, J. Smejkal, \emph{Nucl. Phys.} {\bf A 919} (2013) 46.
\bibitem{gr02} T. Inoue, E. Oset, \emph{Nucl. Phys.} {\bf A 710} (2002) 354.
\bibitem{fix17} A. Fix, O. Kolesnikov, \emph{Phys. Lett.} {\bf B 772} (2017) 663. 
\bibitem{cit1} K. Varga, Y. Suzuki, \emph{Phys. Rev.} {\bf C 52} (1995) 2885. 
\bibitem{mn} D. R. Thompson, M. LeMere, Y. C. Tang, \emph{Nucl. Phys.} {\bf A 286} (1977) 53.
\bibitem{arg} R. B. Wiringa, S. C. Pieper, \emph{Phys. Rev. Lett.} {\bf 89} (2002) 182501.
\bibitem{nc14} P. Navr\'{a}til, E. Caurier, \emph{Phys. Rev.} {\bf C 69} (2004) 014311.
\bibitem{wrw} T. Wass, M. Rho and W. Weise, \emph{Nucl. Phys.} {\bf A 617} (1997) 449.
\bibitem{fg17} E. Friedman, A. Gal, \emph{Nucl. Phys.} {\bf A 959} (2017) 66.   
\bibitem{hm17} J. Hrt\'{a}nkov\'{a}, J. Mare\v{s}, \emph{Phys. Lett.} {\bf B 770} (2017) 342; 
\emph{Phys. Rev.} {\bf C 96} (2017) 015205.
\bibitem{car02} C. Garc\'{\i}a-Recio, T. Inoue, J. Nieves, E. Oset, \emph{Phys. Lett.} {\bf B 550} (2002) 47.
\bibitem{wasa} P. Adlarson et al. (WASA-at-COSY Collaboration), \emph{Nucl. Phys.} {\bf A 959} (2017) 102.
\end{thebibliography}
\end{document}